# Slow Spin Relaxation in Two-Dimensional Electron Systems with Antidots

Yuriy V. Pershin and Vladimir Privman

*Center for Quantum Device Technology,*
*Department of Physics and Department of Electrical and Computer Engineering,*
*Clarkson University, Potsdam, New York 13699-5720, USA*

**Abstract:** We report a Monte Carlo investigation of the effect of a lattice of antidots on spin relaxation in two-dimensional electron systems. The spin relaxation time is calculated as a function of geometrical parameters describing the antidot lattice, namely, the antidot radius and the distance between their centers. It is shown that spin polarization relaxation can be efficiently suppressed by the chaotic spatial motion due to the antidot lattice. This phenomenon offers a new approach to spin coherence manipulation in spintronics devices.

A number of semiconductor devices based on manipulation of electron spin, generally referred to as spintronic devices, have been proposed and simulated [1-21]. Experimental work toward implementation of some of them has been initiated recently [22,23]. In these devices the electron spin control is accomplished primarily by the spin-orbit interactions. Once injected into a semiconductor, the electrons spin polarization will be eventually lost by various relaxation mechanisms. Understanding these mechanisms, as well as development of methods of spin coherence manipulation, are of considerable current interest.

The antidot arrays in semiconductor heterostructures with two-dimensional electron gas (2DEG) have been a model system that allowed study of chaotic classical dynamics in condensed-matter physics [24-39]. The typical spacing of antidots, $a \gtrsim 2000$ Å, is larger than the Fermi wavelength of the 2DEG, which allows to treat the electron spatial motion semiclassically [38]. Various interesting phenomena have been observed in antidot lattices in magnetic fields, including quenching of the Hall effect [26,27], Altshuler-Aronov-Spivak oscillations [28,29], commensurability peaks in magnetoresistance [30-32], and fine oscillations around them [33]. Moreover, this system has been considered as an experimental realization of the theoretical model of Sinai billiard [39]. In this work we report the first investigation of *spin* dynamics in such a system.

We propose to use a two-dimensional electron system, for example, 2DEG in an heterostructure, with a lattice of antidots in spintronic device engineering. In the ideal case, electrons move semiclassically in a plane containing reflecting disks (antidots) of radius $r$, centered at the sites of a square lattice with lattice spacing $a$, as shown in Figure 1(a). Lattice of antidots can be formed when, e.g., a periodic array of holes is etched into the top layers of a semiconductor heterostructure by means of conventional nanofabrication. Based on experimental results, e.g., [40-42], we consider the D'yakonov-Perel' (DP) mechanism [43,44] to be the dominant spin relaxation channel. Using a Monte Carlo simulation scheme originally proposed in [17,18], we calculate the electron spin relaxation time due to the DP mechanism, for varying the spacing $a$ between the antidot centers, the antidot radius $r$, and the strength of the spin-orbit interaction. We have discovered an interesting pattern of dependence of the spin relaxation time on the geometrical parameters of the antidot lattice. These results are presented below. Moreover, we propose to use this system in future spintronic devices, for example, as a new method for coherence control in a spin field-effect transistor.

The DP relaxation results from spin-orbit interactions which cause $\vec{k}$-dependent splitting of the spin states in the conduction band for a wave vector $\vec{k} \neq 0$. This spin splitting can be regarded as an effective magnetic field inducing precession of the electron spin polarization vector, $\vec{S}$, with angular frequency $\vec{\Omega}$. The quantum mechanical evolution of the electron spin polarization vector, defined in a standard way via the single-electron density matrix $\rho$ [45],

$$\vec{S} = Tr(\rho \vec{\sigma}) , \qquad (1)$$

where $\vec{\sigma}$ is the Pauli-matrix vector corresponding to the electron spin, can be described by the equation of motion $d\vec{S}/dt = \vec{\Omega} \times \vec{S}$ [45]. Within the semiclassical approximation, the electrons are treated as classical particles, except that their kinetic energies are determined by the semiconductor energy bands, most commonly in the effective-mass approximation. We assume that the electrons move along trajectories, which are defined by bulk scattering events (scattering on phonons, impurities, etc.) and by scatterings on antidots. Momentum scattering reorients the direction of the precession axis, making the orientation of the effective magnetic field random and trajectory-dependent, thus leading to average spin relaxation (dephasing). Making the trajectory more random/chaotic may actually suppress relaxation, similarly to motional narrowing in NMR [46].

There are two sources of spin-orbit coupling in two-dimensional heterostructures: the inversion asymmetry of the confining potential and the lack of inversion symmetry of the crystal lattice (such as in zinc-blend-lattice semiconductors). The first mechanism yields the Rashba spin-orbit coupling [47],

$$H_R = \alpha \hbar^{-1} \left( \sigma_x p_y - \sigma_y p_x \right) , \qquad (2)$$

where $\alpha$ is a constant, and $\vec{p}$ is the momentum of the electron confined in two-dimensional geometry. The second source of the spin-orbit coupling yields the Dresselhaus interaction [48]. We restrict our consideration to the Rashba spin-orbit coupling, because even in zinc-blend semiconductors it is possible to suppress the Dresselhaus coupling by the appropriate heterostructure growth protocols [49].

The angular frequency corresponding to the Rashba coupling can be expressed as $\vec{\Omega} = \eta \vec{v} \times \hat{z}$, where $\eta = 2\alpha m^* \hbar^{-2}$, $m^*$ is the effective electron mass, $\vec{v}$ is the electron velocity, and the $\hat{z}$ axis is perpendicular to the 2DEG. The spin of a particle moving ballistically over a distance $1/\eta$ will rotate by the angle $\varphi = 1$. The angle of the spin rotation per mean free path, $L_P$, is given by $\Delta \varphi = L_P \eta$. We will assume that at the initial moment of time the spins of the electrons are polarized in $\hat{z}$ direction. We calculate $\langle \vec{S} \rangle$ as a function of time by averaging over an ensemble of electrons. The spin relaxation time is evaluated by fitting the time-dependence of $\langle \vec{S} \rangle$ to an exponential decay. The detailed description of the Monte Carlo simulation method used can be found in [18].

The time-dependence of $\langle \vec{S} \rangle$ was calculated for an ensemble of $10^5$ electrons, for each value of the antidot radius and the lattice spacing. In Figure 2, we plot an example of an electron trajectory obtained in the simulation. It is assumed that the antidot lattice is perfectly reflecting: the electron motion is allowed only in the regions between the antidots. We use the elastic boundary conditions as shown in Figure 1(b). An important property of the electron trajectory exemplified in Figure 2 is that it tends to become chaotic.

The main results of our Monte Carlo simulations are presented in Figures 3 and 4. Figure 3 shows the spin relaxation time, $\tau_s$, extracted from the time-dependence of $\langle \vec{S} \rangle$, as a function of the antidot radius $r$, at fixed selected values of the antidot center spacing, $a$. The electron spin relaxation time as a function of the spacing between the antidot centers, at fixed values of the aspect ratio, $r/a$, is presented in Figure 4. The common feature of all the curves in Figure 3 is that the spin relaxation time increases with decreasing $a$ and with increasing $r$. All the curves in Figure 3 start at the same value at $r = 0$, corresponding to the absence of the antidot lattice. The spin relaxation time is the shortest in this case, because it is determined only by the bulk scattering events. With increasing the antidot radius, the rate of the electron scattering by the antidots increases as well, which results in more frequent random walk-like motion of the polarization vector in the spin-vector space and, consequently, in slower relaxation. The same mechanism explains the increase of the spin relaxation time with decrease of the lattice spacing, observed in Figure 4. It follows from the data shown in Figure 4 that significant (several-fold) increase of the spin relaxation time can be obtained when the distance between the antidot circumferences is of the order of the electron mean free path.

Let us consider the data presented in Figure 3 in detail. Dependence of the spin relaxation time on $r$ can be classified in three different regimes. For small $r$, the dependence is not exponential. Increase of the electron spin relaxation time in this regime is most pronounced for small $a$; see the top curve in Figure 3. Next there follows the regime when the $r$-dependence of the spin relaxation time is approximately exponential, $\tau \sim e^{\gamma(a)(r/a)}$, see the straight line fits on Figure 3. This dependence is valid over almost half of the range of change of the antidot radius, approximately for $0.1 < r/a < 0.35$. The quantity $\gamma(a)$ decreases with increasing $a$. For larger $r$, we observe transition to non-exponential behavior or possibly to an exponential behavior with different slope.

We have also compared these results with the results of a Monte Carlo simulation made with the assumption of "rough" antidots, for which we choose randomly, in $[-\pi/2, \pi/2]$ from the radial direction, the angle of motion of an electron after scattering from an antidot. As illustrated in Figure 3, the spin relaxation time is then only slightly longer than the spin relaxation time with the same system parameters for the reflecting antidots and has almost the same dependence on the antidot radius. This increase in the spin relaxation time likely arises from additional randomization of the electron spatial trajectory by "rough" scattering events.

Spin relaxation control by the antidot lattice can be used in future spintronic devices. Spin polarization can be preserved, and it relaxation rate controlled, by changing the geometrical parameters of the 2DEG. An efficient control over the spin relaxation time can be achieved by an array of circular metal gates located under the 2DEG. The antidot lattice in such a system can be created and controlled by the gate potential. This idea can be used, for example, in engineering of spin field-effect transistors that utilize gate control over the spin relaxation time in 2DEG [16].

In conclusion, we studied relaxation of the electron spins in 2DEG with the antidot lattice. Monte Carlo simulation results indicate that the D'yakonov-Perel' relaxation mechanism in such a system can be efficiently suppressed by the antidot lattice. Spin polarization relaxation time was calculated as a function of the antidot radius and antidot-center lattice spacing. It was observed that in some range of the parameters, the electron spin relaxation time as a function of radius at fixed lattice spacing can be described by an exponential law. While quantitative description of the obtained dependences requires further work, qualitatively the mechanism of suppression of the spin relaxation in 2DEG with antidot lattice can be described as follows. An additional mechanism of scattering of the electrons by the antidots, and, correspondingly, the reduction of the electron mean free path, and the chaotic nature of the spatial trajectory, lead to rapid changes in the effective spin-orbit "magnetic field" experienced by the electron spin. Therefore, the spin rotations become random-walk-like. For each electron, then, the overall spin drift from the original polarization direction is actually reduced. Since in our semiclassical description the DP relaxation results from averaging over an ensemble of electrons, it is actually suppressed when each electron's spin drifts less from the original direction. In summary, the considered experimentally realizable system offers new ways to achieve long electron spin relaxation times in spintronics devices.

We gratefully acknowledge helpful discussions with Drs. L. Fedichkin and S. Saikin. This research was supported by the National Security Agency and Advanced Research and Development Activity under Army Research Office contract DAAD-19-02-1-0035, and by the National Science Foundation, grant DMR-0121146.

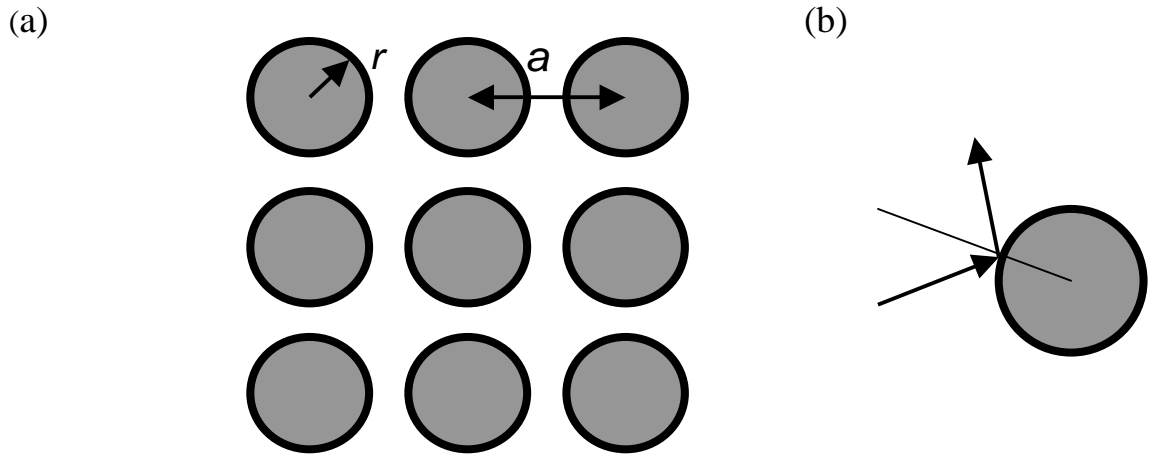

**Figure 1.** (a) The antidot lattice. (b) Elastic reflection of an electron from an antidot.

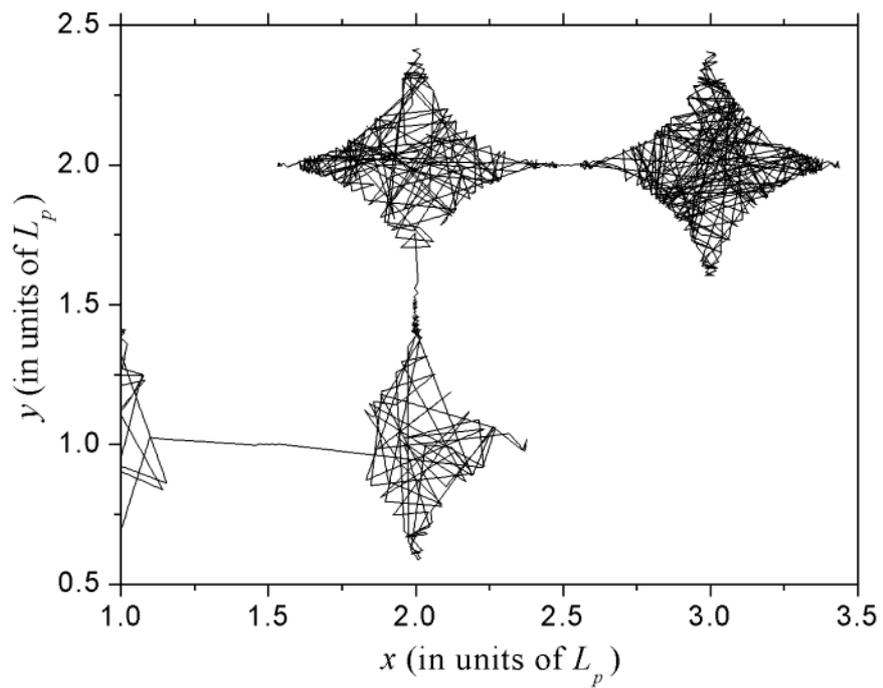

**Figure 2.** Example of an electron trajectory when antidots are almost touching each other.

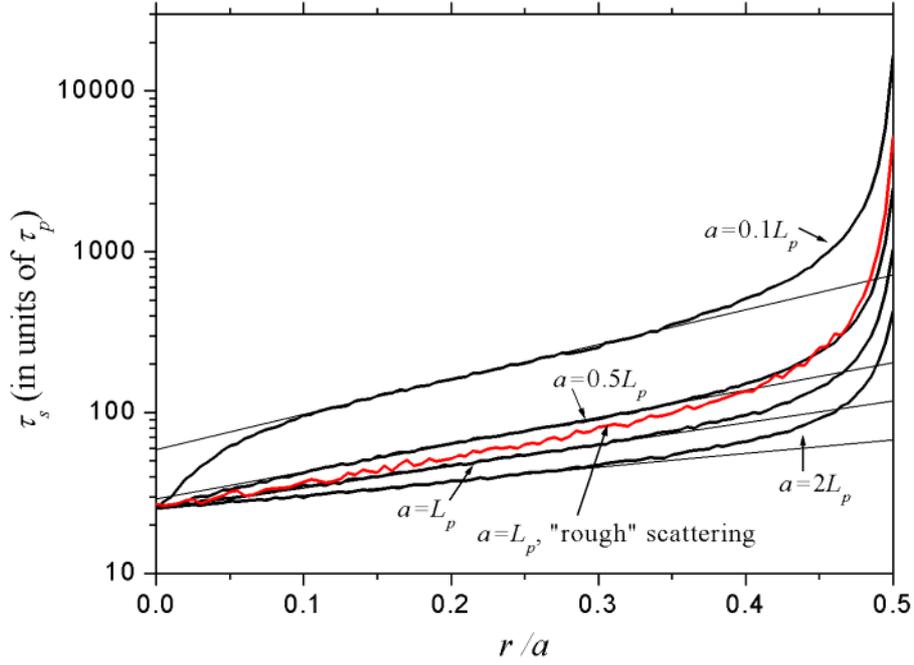

**Figure 3.** Electron spin relaxation time, $\tau_s$, as a function of the antidot radius, for different spacing between the antidots, with $\eta L_p = 0.2$. The straight lines are the fitted exponentials; $\tau_p$ is the momentum relaxation time.

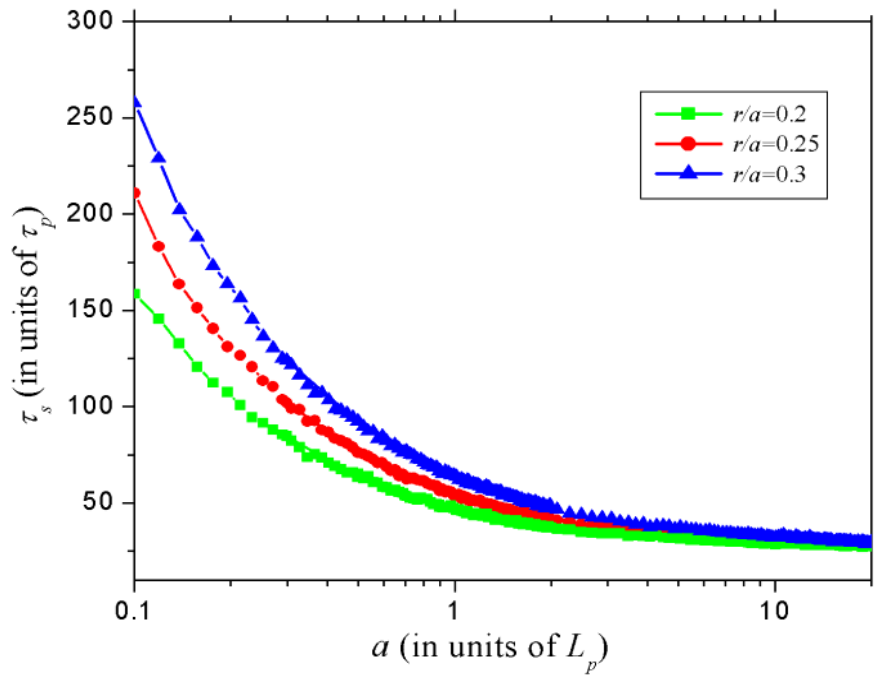

**Figure 4.** Relaxation time at fixed $r/a$, as a function of the spacing between the antidot centers.